\documentclass[newstyle,preprint]{rmaa}
\usepackage{rmaacite}

\newcommand{\ii}{\'{\i}}

\newcommand{\upar}{\hbox{$\bar U$}}

\newcommand{\tsq}{\hbox{$t^2$}}
\newcommand{\tobs}{\hbox{$t^2_{obs.}$}}
\newcommand{\tref}{\hbox{$t^2_{steady}$}}
\newcommand{\tasym}{\hbox{$\left<t^2\right>$}}
\newcommand{\tasymo}{\hbox{$\left<t^2{\rm (O^{+2})}\right>$}}
\newcommand{\tasymh}{\hbox{$\left<t^2{\rm (H^{+})}\right>$}}

\newcommand{\teq}{\hbox{$T_{eq}$}}
\newcommand{\tom}{\hbox{$\left<\bar T_0\right>$}}
\newcommand{\tmean}{\hbox{$\bar T_0$}}
\newcommand{\trec}{\hbox{$\tau_{rec}$}}
\newcommand{\tcool}{\hbox{$\tau_{cool}$}}
\newcommand{\tcros}{\hbox{$\tau_{cross}$}}

\newcommand{\ha}{\hbox{H$\alpha$}}

\newcommand{\nubar}{\hbox{$\overline{\nu}$}}
\newcommand{\abar}{$\overline{a}$}
\newcommand{\nuzero}{$\nu_0$}

\newcommand{\nH}{n$_{\mbox{\scriptsize H}}$}
\newcommand{\ygn}{{\mbox{\sc yguana}}}
\newcommand{\ygz}{{\mbox{\sc yguazu}}}
\newcommand{\map}{{\mbox{\sc mappings~i}c}}
\newcommand{\Rin}{R$_{\mbox{\scriptsize in}}$}
\newcommand{\Rout}{R$_{\mbox{\scriptsize out}}$}
\newcommand{\thetac}{\hbox{$\theta^1$\,C~Ori}}

\newcommand{\ten}[1] {10$^{#1}$}
\newcommand{\cmc}{cm$^{-3}$}

\newcommand{\hi}{\hbox{H\,{\sc i}}}
\newcommand{\hii}{\hbox{H\,{\sc ii}}}

\newcommand{\hp}{\hbox{H$^{\rm +}$}}

\newcommand{\opp}{\hbox{O$^{\rm +2}$}}

\newcommand{\heii}{\hbox{He\,{\sc ii}}}

\newcommand{\ciiiw}{\hbox{C\,{\sc iii}]$\lambda\lambda $1909}}
\newcommand{\oiii}{\hbox{[O\,{\sc iii}]}}
\newcommand{\oiiiw}{\hbox{[O\,{\sc iii}]$\lambda $5007}}
\newcommand{\oiiitw}{\hbox{[O\,{\sc iii}]$\lambda $4363}}

\title{Relation between source and temperature fluctuations in
photoionized nebulae}

\author{Luc Binette \altaffilmark{1}, Pierre Ferruit \altaffilmark{2},
Wolfgang Steffen \altaffilmark{1} and 
Alejandro Raga \altaffilmark{3}
}

\altaffiltext{1} {Instituto de Astronom\'{\i}a, Universidad Nacional
Aut\'onoma de M\'exico  }
\altaffiltext{2} {CRAL -- Observatoire de Lyon}
\altaffiltext{3} {Instituto de Ciencias Nucleares, Universidad Nacional
Aut\'onoma de M\'exico  }

\fulladdresses{
\item L. Binette:  Instituto de Astronom\'{\i}a, UNAM, Apdo. Postal
70-264, 04510 M\'exico, D. F., M\'exico  
(binette@astroscu.unam.mx). 
\item P. Ferruit: CRAL - Observatoire de Lyon, 9 av. Charles Andr\'e,
F-69561 Saint-Genis-Laval Cedex, France.
\item W. Steffen: Instituto de Astronom\'{\i}a, Apdo. Postal 877,
22830 Ensenada, B. C., M\'exico.
\item A. Raga: Instituto de Ciencias Nucleares, UNAM, Apdo. Postal 70-543,
04510  M\'exico, D. F., M\'exico.
}

\shortauthor{Binette et al.}
\shorttitle{Time-dependent photoionization}

   \abstract{ The magnitude of the temperature fluctuations (\tsq)
   required to explain the observed inconsistencies between
   metallicities inferred from recombination lines and from forbidden
   lines cannot be attained by steady-state equilibrium
   photoionization models. If on the other hand the nebular ionizing
   source was variable, the temperature fluctuations \tsq\ would be
   significantly larger. We investigate the time-dependent response of
   the nebular ionization and temperature structure when photoionized
   by a periodically varying source. We study how the asymptotic mean
   value, \tasym, behaves as a function of the period or amplitude of
   the source variability. We find that the temperature fluctuations
   occur only in the outer section of the nebula, close to the
   ionization front, within a zone corresponding to 8--20\% of
   the ionized layer's thickness. We conclude that the amplitude of
   the exciting star variations required to achieve a $\tasym = 0.025$
   (as in the Orion nebula) is unacceptably large. Source variability
   is therefore not a viable mechanism to explain the observed values
   of \tsq. We reach a similar conclusion from studies of the temporal
   variability resulting from intermittent shadows behind opaque
   condensations. We find that photoionized nebulae are on average
   less massive but somewhat hotter in the case of cyclicly variable
   ionizing sources.

}

\resumen{ La magnitud de las fluctuaciones de temperatura (\tsq) que
se necesitan para explicar las incoherencias observadas entre las
metalicidades inferidas por las l\ii neas de recombinaci\'on o por las
l\ii neas prohibidas, no pueden alcanzarse con modelos al equilibrio y
estacionarios en el tiempo. Por otra parte, si la fuente ionizante
fuera variable, las fluctuaciones de temperaturas \tsq\ ser\ii an
mucho mayores. Investigamos la respuesta temporal de la estructura en
ionizaci\'on y temperatura de una nebulosa fotoionizada por una fuenta
variable peri\'odica. Estudiamos como el valor medio asimpt\'otico,
\tasym, se comporta en funci\'on del peri\'odo y amplitud de las
variaciones de la fuente. Encontramos que las fluctuaciones de
temperatura se producen solamente en la parte externa de la nebulosa,
cerca del frente de ionizaci\'on, dentro de un espesor geom\'etrico
correspondiente al 8-20\% del tama\~no de la capa ionizada. Concluimos
que la amplitud de variaci\'on de la estrella excitadora que se requiere para
conseguir un $\tasym = 0.025$ (como en la nebulosa de Orion) es
excesivamente grande. La variabilidad de la fuente ionizante no es por
lo tanto un mecanismo viable para explicar los valores observados de
\tsq. Llegamos a conclusiones similares cuando estudiamos la
variabilidad temporal que resulta de sombras intermitentes detr\'as de
condensaciones de gas opacas. Encontramos que nebulosas fotoionizadas
son en promedio menos masivas pero algo m\'as calientes en el caso de
fuentes variables c\ii clicas.

}
\nonstopmode
\begin{document}
\maketitle


\section{Introduction} \label{SectionIntroduction} 
The presence of temperature fluctuations in photoionized nebulae has
been a matter of debate since the pioneering work of  
\scite{peim67}. Observational evidence
has since accumulated in favor of his analysis of the problematics of
nebular temperatures. For instance, the temperatures of
\hii\ regions and planetary nebulae are observed to be significantly
lower when derived using recombination lines rather than from
forbidden line ratios (see review by \pcite{peim95}).

The work of \scite{king95} and  
\scite{per97} has shown that the amplitude of the fluctuations in
hydrostatic photoionization calculations are much smaller than
required to explain the observed differences between forbidden and
recombination lines temperatures. Possible causes for the fluctuations
include metallicity inhomogeneities (\pcite{tor90}; \pcite{king98}; 
\pcite{liu00}), photoionization by small dust grains
(\pcite{sta01}), and shock heating due to stellar wind or even
supernovae in the case of giant \hii\ regions (\pcite{lur99}).  The
inclusion of temperature fluctuations in a theoretical framework has
so far been tentative (\pcite{lur01}; \pcite{bin00}; \pcite{bin01})
given our lack of knowledge about their possible cause.

In this Paper, we investigate how temporal variability in ionization
structure due to an intrinsically variable ionizing source or to 
intermittent shadows behind opaque condensations can affect the
temperature structure and induce substantial temperature
fluctuations. We also address the question whether a
variable ionizing continuum with a fixed duty cycle would lead to a  
nebula that is on average warmer and less massive.

\section{Calculations} \label{sec:calcul} 

\subsection{Intrinsic variability of the ionizing source } \label{sec:var}

Resolution of the ionizing radiation transfer in the case of a
variable source implies a sufficiently high temporal resolution in
order to follow the progression of ionizing photon fronts across the
nebula. The atomic timescales on the other hand are much longer, and a
lot of computation time is spent in the actual transfer while not much
is occurring in terms of changes in ionization or temperature.  For
this reason, in the building of the new code \ygn\ we simplified the
atomic physics in order to derive the desired results within
reasonable computation efforts. Our goal was to give precedence to a
rigorous treatment of the transfer, and not to detailed atomic
physics. The Courant condition, for instance, was rigorously satisfied
in all the calculations (Eq. \ref{EquationStepSize} in Appendix
\ref{sec:cour}).

One of the simplifications introduced in \ygn\ is that only Hydrogen
is considered in the integration of the nebular opacity, which is a
valid approximation for \hii\ regions. Furthermore, we consider a
monochromatic transfer with all ionizing photons having the same
energy. This simplification implies that the hardening of the ionizing
radiation with depth is not considered. The expected shallow radial
temperature gradient will therefore be absent from our calculations.
This is a minor shortcoming of \ygn, since the level of temperature
fluctuations caused by non-equilibrium ionization largely exceed that
produced by the temperature gradient alone (c.f.
\S \ref{sec:steady}).  In effect, non-equilibrium photoionization
within the IF, which we compute accurately, far exceeds the larger
incdreased due to continuum hardening at large depths.  The various
effects introduced by UV source variability on a photoinized nebula
will all be accurately tracked and made obvious by comparing our
time-dependent nebula to either the time-averaged nebula or to the
steady-state constant-UV reference nebula.

The cooling by metals was approximated in \ygn\ by a fit we made to
the cooling function using \map\ (\pcite{fer97}) [c.f.
Appendix~\ref{AppendixEquations}]. Collisional excitation and
ionization of \hi\ (and their effect on cooling) are calculated
explicitly, since they can become important within the ionization front
(IF). The time-dependent ionization balance of H, the photoheating
coefficients, and the equation of state are described in
Appendix~\ref{AppendixEquations} while the practical implementation of
the time-dependent transfer equations are presented in
Appendix~\ref{AppendixCode}. Our main goal is to study the effect on
the temperature fluctuations measured {\it
\`a la} Peimbert as a result of having a variable ionizing source. Our 
model assumes an hydrostatic distribution of gas.

\subsection{Mean temperature \tmean, \tsq\ and  \upar  } \label{sec:def}

Following \scite{peim67}, we define the mean nebular temperature,
\tmean, as follows

\begin{equation}
\bar T_0(\hp)=\frac {\int_V n_e n_{H^{+}} T dV}{\int_V n_e n_{H^{+}} dV} \; , \label{eq:to}
\end{equation}

\noindent where
$n_e$ is the electron density, $n_{H^{+}}$ the \hii\ density, $T$
the electron temperature and $V$ the  volume over which the
integration is carried out. The rms amplitude $t$ of the temperature
fluctuations is given by

\begin{equation}
t^2(\hp) = \frac {\int_V n_e n_{H^{+}} (T - \bar T_0)^2 dV} 
{\bar T_0^2\int_V n_e n_{H^{+}} dV} \; . \label{eq:tsq}
\end{equation}
We have replaced $n_{H^+}$ by $n_e$ in the above expressions, since we
do not consider explicitly He in the calculations. $n$ will denote the 
total H density.

We cannot compute \tsq\ for other ions, since we consider in detail
only the ionization of H. We will drop  the label $\hp$ in
the following discussion unless required by the context. 

In the case of a variable source, both global quantities \tsq\ and
$T_0$ vary with time.  The above expressions define therefore
instantaneous values for a particular time in the source temporal
evolution. Even if the transfer of information concerning $n_{H^{+}}$,
$n_{H^{0}}$ or $T$ from any position inside the nebula has a finite
speed, we do not find necessary to consider this effect explicitly,
since we are not considering any particular external
observer. Furthermore, our derivation of various characteristic
quantities (labeled ``asymptotic'') will be done by integrating over a
few full periods of the source variability. This has the effect of
washing out any phase differences introduced by placing a real
observer at any particular external location.

Let us define further useful quantities to be used later. One is the
varying ionizing source photon luminosity, $Q_H$. Time-averaged
quantities will carry brackets $\left<\right>$ such as in
$\left<Q_H\right>$, the time-averaged photon luminosity. Since we
adopt a spherical geometry and consider the source point-like, the
global ionization parameter is defined as
\begin{equation}
\label{eq:upar}
	    \begin{array}{ll}
\bar U= {\frac {\int_V \epsilon^2 n_e^2  {\frac {\varphi}{c \, n}}
\, dV} {{\int_V \epsilon^2 n_e^2}\, dV}} 

= {{\int^{R_s}_{r_0} \epsilon^2 n_e^2  {\frac {\left<Q_H\right>}{4\pi
r^2 c \, n}} \, 4 \pi r^2 \, dr} \over {{\int^{R_s}_{r_0} \epsilon^2 n_e^2}
4 \pi r^2 \,dr} }  
	    \end{array}
\end{equation}
where $\varphi$ is the local ionizing flux (disregarding opacity),
$\tau_H$ the opacity due to H photoionization, $R_s$ the outer nebular
radius, $r$ the distance from the central source, $r_0$ the radius of
the inner cavity devoid of gas, $\epsilon$ the volume filling factor,
and $c$ the speed of light. (This definition contemplates the
possibility that  $\epsilon$ and the total density $n$  may vary with radius.)


\subsection{Results for periodical sources (Model A)} \label{sec:resu}

Our main goal is to derive the characteristic \tasym\ of a nebula
submitted to a variable ionizing source. The general problem is
excessively vast since there are numerous scenarios about possible
variability behaviors.  The perspective taken in this Paper is that of
a Fourier analysis in which we restrict ourselves to the exploration of
periodical sources and analyze the nebular response as a function of
the frequency and amplitude of the source variability.  In concrete terms, we have narrowed the problem to deriving the {\it asymptotic} \tasym\ for a source varying at a predetermined frequency. We define the asymptotic
\tasym\ as the averaged value over at least three full periods of the
source.  This has the advantage of avoiding a definition that
depended on the initial conditions or on a particular ill-defined
moment in the source history. Typically, calculations will extend to
only five periods since we found that asymptotic values in most cases
do not change by adding more cycles.

   \begin{figure*}
   \resizebox{12cm}{!}{\includegraphics{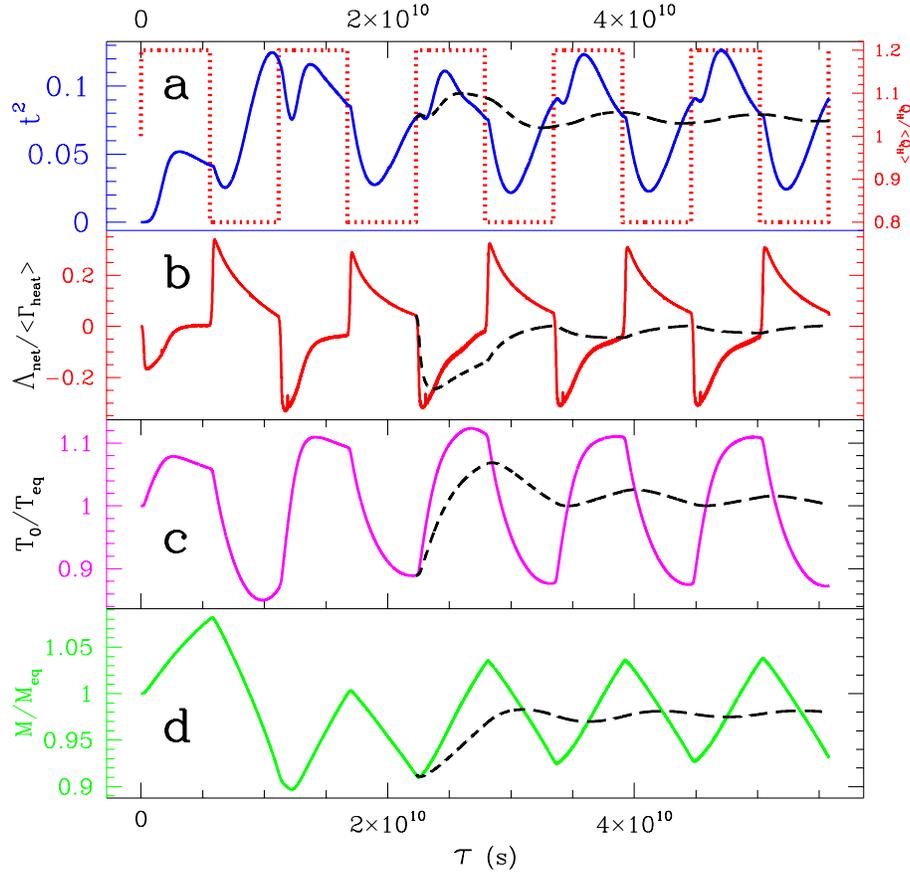}} \hfill
   \parbox[b]{55mm}{

\caption{All solid lines: behavior of nebular volume
weighted quantities as a function of time $\tau$ for Model
A. Panel~{\it a}:
\tsq, Panel~{\it b}: net cooling relative to the mean heating energy
radiated by the source $\Lambda_{net}/\left<\Gamma_{heat}\right>$
[with $\left<\Gamma_{heat}\right> = h (\nubar -\nu_0)
\left<Q_H\right>$], Panel~{\it c}: the mean electron temperature
$\tmean/\teq$, and Panel~{\it d}: the mass of ionized gas $M/M_{eq}$. In
each panel, the long-dashed line across the last three cycles
represents the running mean of the plotted quantity. The asymptotic
values referred to in the text (e.g. \tom, \tasym) correspond to the
rightmost value along the long-dashed line.  The {\it square-wave}
dotted-line in Panel~{\it a} is the normalized photon luminosity of the
ionizing source as a function of time, $Q_H(\tau)/\left<Q_H\right>$
(to be read on the upper right $y$-axis). The variability is
characterized by an amplitude 
$A=0.2$, a period   $\Pi = \trec$ and a duty cycle $\Delta =
0.5$. \label{fig:oscil}}}
\end{figure*}

A varying source generate two types of response in the nebular
structure: a progressing IF when the source increases in intensity and
a recombination front (RF) when the source decreases. These two phases
are not symmetric. For instance, an IF has a finite velocity given by
$V_{IF} \simeq \varphi/n_{H^0} < c$, where $\varphi$ is the ionizing
photon flux. Let us define $y$ as the neutral fraction of H. The
ionization fraction ($1-y$) within the IF can increase at a rate,
which is much shorter than the recombination timescale. An RF on the
other hand propagates at the speed $c$, since a RF is initiated
following propagation of the signal that the source is decreasing
(although the neutral H fraction $y$ within the front will increase
only on timescales given by recombination).

To illustrate these two phases, using \ygn\ we will study in detail
the case of source varying like a square-wave\footnote{We will use
primes to denote models that assume a {\it sine-wave} (rather than
{\it square-wave}) variability, e.g. Model A$^{\prime\prime}$ in \S
\ref{sec:perio}.}  with a period $\Pi$ equal to the recombination
timescale $\trec = (\alpha_B n_e)^{-1}$, where $\alpha_B$ is the
recombination coefficient rate to excited states of H. The amplitude
of the variability is $\pm 20\%$ ($A=0.20$) and the duty cycle is
$\Delta=0.5$. The advantage of the square-wave is that it will more
clearly reveal the changes caused by either type of fronts on certain
nebular quantities.  We will assume an ionizing photon luminosity of
$\left<Q_H\right> =10^{51} {\rm s^{-1}}$ and a nebula of constant
density $n=200$\,\cmc\ with a volume filling factor of $\epsilon =
0.01$ and an inner cavity $r_0$ surrounding the source of 30\% of the
estimated Str\"omgren radius, that is, of size $r_0 = 4.0 \times
10^{19}\,{\rm cm}$.  The initial physical conditions of the gas were
given by the usual condition of steady-state thermal and ionization
equilibrium, throughout the whole nebula, for a source luminosity
$\left<Q_H\right>$. The equilibrium temperature within the nebula
turns out to be $\teq = 5260$\,K and the recombination and cooling
timescales within the fully ionized part of the nebula are $\trec =
1.1 \times 10^{10}$\,s and $\tcool = 1.6 \times 10^9$\,s, respectively
(the cooling timescale is 7 times shorter than \trec).  The ionization
parameter (Equation~\ref{eq:upar}) of the initial equilibrium model is
$\upar = 0.0022$. This set of parameters defines our Model~A.

The results of the time-dependent calculations as a function of time
$\tau$ for the first five periods are shown in
Fig.~\ref{fig:oscil}. The dotted line in Panel~{\it a} illustrates the
behavior of the ionizing flux relative to the mean value (to be read
on the upper right $y$-axis of  Panel~{\it a}). In all the panels, the
continuous line corresponds to the behavior of the instantaneous values
integrated over the whole nebula while the long-dashed line represents
the cumulative running mean, which was calculated at the onset of the
third variability cycle. We define asymptotic values as the last value
of the running mean (after the 5th cycle is completed). The impact on
the nebula of the progression of the IF or the recession of the RF are
clearly visible in Panel~{\it d}, which is a plot of the mass of ionized gas
normalized to the initial equilibrium value. Notice that the total
ionized mass of the nebula somewhat shrinks with time.  Panel~{\it c} shows
the behavior of the average nebular temperature with time,
$\tmean(\tau)$, normalized again to the initial equilibrium value
\teq. Panel~{\it b} shows the behavior of the net integrated cooling rate of
the nebula normalized to the the total heating available due to
absorption of all ionizing photons (this ratio cannot exceed the
interval $\pm 1$). With steady-state equilibrium, the net cooling is
zero. Panel~{\it b} reveals how the nebula as a whole heats up during the IF
phase while it cools down during the RF phase. Finally, Panel~{\it a}
illustrates the behavior of \tsq\ as a function of time. Its
(asymptotic) average value is $\tasym = 0.074$.

\subsection{Missing the temperature gradient} \label{sec:steady}

Not considering the hardening of the UV radiation with depth results
in an isothermal nebula lacking the usual outward gradient in
$T_{eq}$.  This has the advantage that \tsq\ computed with \ygn\ is
determined by source variability alone. The effect of missing the
shallow $T_{eq}$ gradient is negligible in most cases. In effect,
using \map, we find that the constant-UV steady-state Model A is
characterized\footnote{The values of $\tsq(\opp)$ computed by \map\ in
the constant-UV case for Models A and B are 0.0021 and
0.0047,respectively.} by a $\tref=0.0026$, which is generally very
small in comparison with the values derived for the variable sources
studied below.  The true \tsq, which would include UV hardening,
cannot be far off from the following estimate $t^2_{true} \approx
\tref + \tasym$, where \tasym\ corresponds to the time-averaged
fluctuation amplitudes computed using \ygn. We recall that
$\tasym = 0.074$ for Model A with $A=0.2$.

\subsection{The internal nebular structure} \label{sec:struc}

\begin{figure*}
   \resizebox{12cm}{!}{\includegraphics{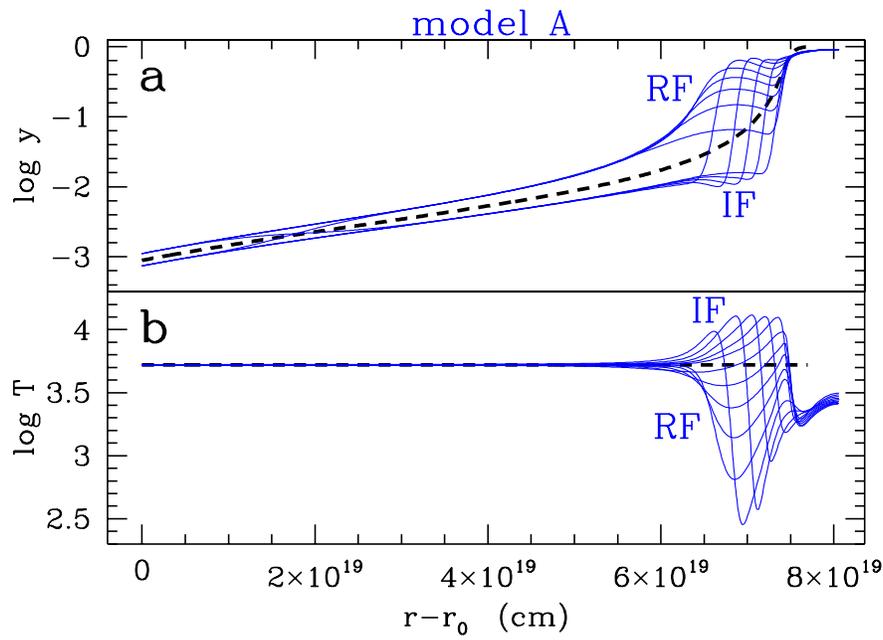}} \hfill
   \parbox[b]{55mm}{ \caption{Internal structure of the nebula as a
   function of geometrical depth for Model A (c.f. \S~\ref{sec:resu})
   assuming a {\it square-wave} variability of amplitude
   $A=0.20$. Panel~{\it a}: the neutral fraction $y$ of H, Panel~{\it b}: the
   local temperature $T$. The long-dashed line represents the initial
   equilibrium model. The 10 solid lines correspond to the ionization
   and thermal structure at equally spaced time intervals during the
   last full cycle shown in Fig.~\ref{fig:oscil}a.
\label{fig:struca}}}  
\end{figure*}

The changes taking place within the spherical nebula of Model A occur
mostly within the external parts, that is within the outer 25\%, which
corresponds nonetheless to half the photoionized volume. This becomes
apparent in Fig.~\ref{fig:struca} where the H neutral fraction, $y$,
and the local temperature are plotted as a function of nebular
geometrical depth $r-r_0$. The long-dashed line represents the initial
equilibrium model. The lack of any slope in the equilibrium
temperature curve (dashed-line) in Panel~{\it b} is the result of not
considering the hardening of ionization radiation with depth (the
transfer in \ygn\ is monochromatic, but see \S \ref{sec:steady}). The 10
solid lines correspond to the ionization and thermal structure at
equally spaced time intervals during the last (fifth) variability
cycle (see last cycle in Fig.~\ref{fig:oscil}a). The IF phase
corresponds to the 5 curves below the dashed-line in Panel~{\it a} while the
other 5 curves correspond to the RF phase. During the RF phase, the
neutral fraction lies above the dashed-line equilibrium curve (except
near the transition to neutral gas, where $y\sim 1$). The progression
of the IF is characterized, on the other hand, by a sharp temperature
pulse above the equilibrium value, which is spatially resolved in
\ygn\ (see Fig.~\ref{fig:struca}b). This IF is sufficiently hot to
cause a rise in the mean temperature (averaged over the whole nebula)
as seen in Fig.~\ref{fig:oscil}c.  Because the selected source period
is rather long, being equal to \trec, most of the inner nebula has
time to adjust its ionization while the source varies. This is shown
by the coincidence of most curves for geometrical depths $r-r_0 < 6
\times 10^{19} \, {\rm cm}$, as they lie (and superimpose) either
below or above the equilibrium $y$ dashed-line model in
Fig.~\ref{fig:struca}a.

   \begin{figure*}
   \resizebox{12cm}{!}{\includegraphics{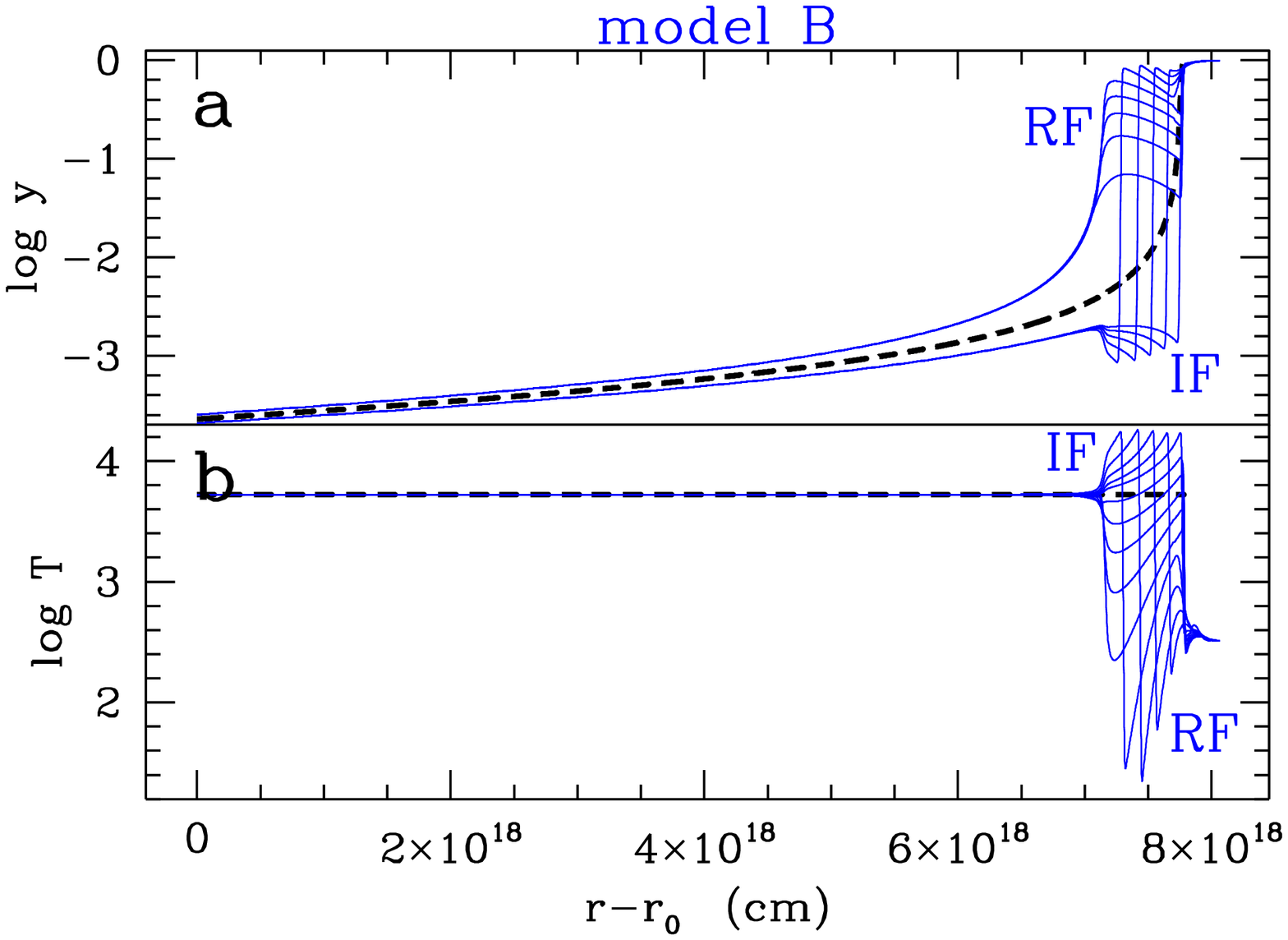}} \hfill
   \parbox[b]{55mm}{ \caption{Internal structure of the nebula as a
   function of geometrical depth for Model B (c.f. \S~\ref{sec:upar})
   assuming a {\it square-wave} variability of amplitude
   $A=0.10$. Panel~{\it a}: the neutral fraction $y$ of H, Panel~{\it b}: the
   local temperature $T$. The long-dashed line represents the initial
   equilibrium model. The 10 solid lines correspond to the ionization
   and thermal structure at equally spaced time intervals during the
   last full cycle.
\label{fig:strucb}}}  
\end{figure*}

Interestingly, the temporal changes in the neutral fraction $y$ with
time for depths $r-r_0 \le 5 \times 10^{19} \, {\rm cm}$ does not cause
any appreciable changes in the gas temperature (see Panel~{\it b}). The
reason is that during the IF or RF phase, the neutral fraction $y$ in
that zone evolves quickly toward the appropriate equilibrium (but
small) value $y_{eq}$ over a timescale of order $\sim 0.5 y_{eq}
\trec$, which is much shorter than \trec\ (because $y_{eq} < 0.01$ in
that zone).  Hence, non-equilibrium heating is too short-lived in that
inner zone for the temperature to change appreciably.

\subsection{Varying the ionization parameter \upar\ (Model B) }
\label{sec:upar}

The work of \scite{camp88} indicates that the ionization
parameter\footnote{The definition of \upar\ of \scite{camp88} is
slightly different, which causes the values quoted by her to be 50\%
higher than those one would derive using eq. (\ref{eq:upar})} in
\hii\ galaxies  range from 0.0014 to 0.025. Hence Model A ($\upar = 0.0022$) 
is probably representative of nebulae\footnote{Models of different
densities, filling factors and source luminosities but whose product
of $n \times \epsilon^2 \times\left<Q_H\right>$ is the same as well as
the ratio $r_0/R_S^0$, where $R_S^0$ is the Str\"omgren radius in the
cavity-less case ($R_S^0 = [{3}\left<Q_H\right>/4 \pi n^2\epsilon \bar
\alpha_B]^{1/3} $), are homologous and can therefore be expected to
yield the same \tasym\ values under similar variability conditions
[i.e. variations of similar amplitude $A$ and period $\Pi$ (in \trec\
units)].} at the low \upar\ end. We have explored the behavior of
\tasym\ when  the ionization parameter is varied. 
This was done by setting the filling factor $\epsilon$ to unity and
computing models which were characterized by different cavity
sizes. By varying $r_0$ in the interval $2.5 \times 10^{19} \le r_0
\le 1.5 \times 10^{20} $\, cm, we explored the following range
0.018--0.0006 in \upar\ (we assumed a source variability of $A=0.05$
with the other parameters such as $n$, $\Delta$, $\Pi$,
$\left<Q_H\right>$ the same as in Model A). What we found was that
\tasym\ increased approximately as $\sqrt{\upar}$ within the above range.

We also studied the properties of nebulae that reflect more closely
the geometrical particularities of the Orion nebulae, that is a model
consisting of a thin sheet of ionized gas at a much higher ionization
parameter. For instance, the Orion model of \scite{bf91} is
characterized by a slab geometry at an ionization parameter of
$0.03$. We reproduced similar characteristics by using $r_0 = 2
\times 10^{19}$ \, cm and by setting the filling factor 
$\epsilon$ to unity. With $\left<Q_H\right> =10^{51} {\rm s^{-1}}$ and
a density $n=200$\,\cmc, the thickness of the ionized
layer turns out to be $7.1 \times 10^{18}$ \, cm, that is 26\% only of the final
Str\"omgren radius. The ionization parameter (Equation~\ref{eq:upar})
of the initial equilibrium model is $\upar = 0.024$. This set of
parameters defines our Model B.

In Fig. \ref{fig:strucb}, we show the internal temperature and
ionization structure of Model B for a variability amplitude of
$A=0.10$. This model is characterized\footnote{Using \map, we find
that the constant-UV steady-state Model B is characterized by
$\tref=0.01$ (see \S \ref{sec:steady}).} by an asymptotic \tasym\ of
0.053. Overall, Models B and A share many similarities. For instance,
the bulk of the variability in $T$ or $y$ occurs near the outer IF.

   \begin{figure}
   \resizebox{\hsize}{!}{\includegraphics{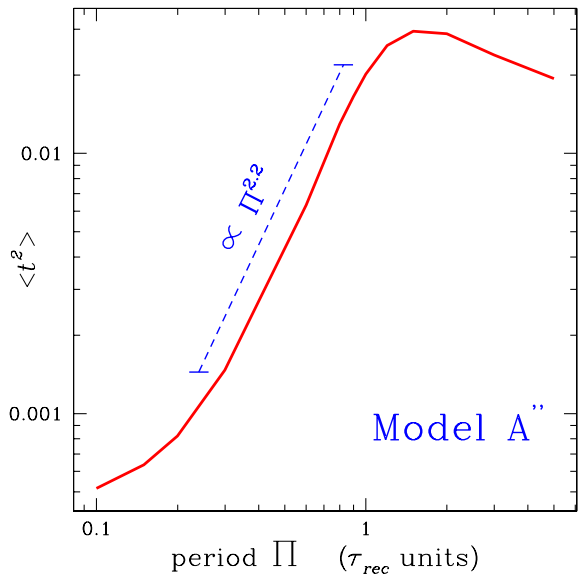}}
   \caption{Behavior of the asymptotic \tasym\ as a function of the
   period $\Pi$ for a {\it sine-wave} variability of amplitude $A=0.1$
   (Model A$^{\prime\prime}$). In the interval $0.2 \le \Pi \le 1$,
   \tasym\ increases steeply as $\Pi^{2.2}$.}  \label{fig:perio}
   \end{figure}

\subsection{Sensitivity to the period $\Pi$} \label{sec:perio}

Using a square wave variability type, we have illustrated the
principal effects taking place within a nebula submitted to a
periodical ionizing source. Two very important parameters affect the
nebular response and the behavior of \tasym: the variability period
$\Pi$ and the amplitude of the variations $A$. We will analyze in
turns the role played by each, adopting the same parameters as in
Model~A but with a sine-wave variability function ($1+ A\sin
2\pi\tau/\Pi$) rather than a square wave. We will refer to this
modified model as Model A$^{\prime\prime}$. We have verified that
sine-wave models\footnote{The values of \tasym\ for models
A$^{\prime\prime}$  and B$^{\prime\prime}$   in the
{\it sine-wave case} with $\Pi = 1 \trec$ and $A=0.1$ are 0.023 and
0.039, respectively. For $A=0.2$, it is 0.055 and 0.078, respectively.}
with amplitude $1.4 A$ closely match the \tasym\ from square-wave
models with amplitude $A$.  All periods will be expressed in units of
the recombination time, which is the natural unit for the problem at
hand.

Our results for Model A$^{\prime\prime}$, assuming an amplitude $A=0.1$,
show a very steep dependence with frequency as illustrated in
Fig.~\ref{fig:perio}. \tasym\ increases as $\propto \Pi^{2.2}$ up to
$\Pi \simeq 1 \trec$ and then peaks near 1.6\trec. The short period
regime becomes progressively ill-defined in our calculations. For
instance, the asymptotic value of \tasym\ at $\Pi = 0.1$ requires up
to 10 cycles, because the mean values keep evolving well beyond 5
cycles.  If radiation hardening at large depths had been considered,
\tasym\ would lie above the  $\tref=0.0026$ floor computed by \map\
(see \S \ref{sec:steady}). This would alter  the curve in Fig.~\ref{fig:perio}
only in the short period domain $\Pi \la 0.4$. 

The main conclusion is that a nebula acts as a low-pass filter with
only the relatively `slow' source variations causing important
temperature fluctuations.  Clearly, $\Pi \ga 0.7$ is the frequency
domain that favors the largest values of \tasym.

\subsection{Sensitivity to the amplitude $A$} \label{sec:ampl}

\begin{figure}
\resizebox{\hsize}{!}{\includegraphics{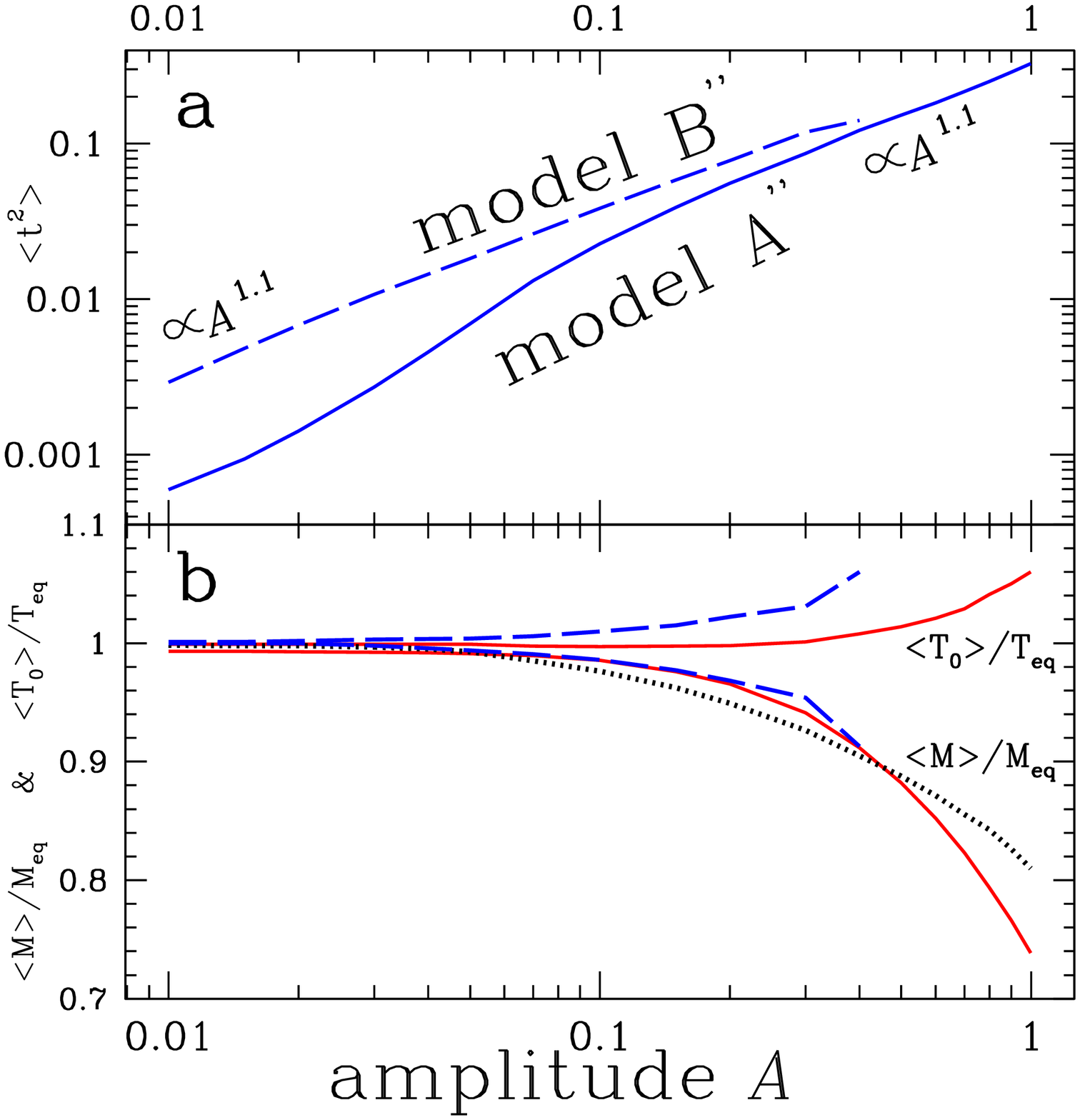}}
\caption{Panel~{\it a}: behavior of \tasym\ as a function of the
amplitude of the {\it sine-wave} variations of $Q_H$. Panel~{\it b}:
behavior of $\tom/\teq$ (curves $\ge 1$) and $\left<M\right>/M_{eq}$
(curves $\le 1$) as a function of the amplitude of the {\it sine-wave}
variations. Model A$^{\prime\prime}$ and B$^{\prime\prime}$ are
represented using solid and dashed lines, respectively. The dotted
line represents the equation $(1+0.7863\tasym)^{-1.177} \tom/\teq$
(see \S~\ref{sec:ampl}) using values from Model
A$^{\prime\prime}$. Model B$^{\prime\prime}$ was not computed beyond
$A=0.4$.}
\label{fig:ampl} 
\end{figure}

Let us now vary the amplitude (up to the maximum, which is $A=1$ in
the case of a sine-wave), adopting the same sine-wave variability with
$\Pi = 1$. The results are shown in Fig.~\ref{fig:ampl}a. The larger
the amplitude, the higher \tasym, as one would expect. In the case of
Model A$^{\prime\prime}$, there is, first a regime of faster increase
($\propto A^{1.8}$) of \tasym, followed by a slower increase regime
($\propto A^{1.1}$) above $A=0.1$.  For Model B$^{\prime\prime}$, the
increase is $\propto A^{1.1}$ everywhere.  The basic result from this
plot is that an observed $\tobs \simeq 0.02$ would require a
substantial variability amplitude of 0.09 and 0.05 from the source,
for Models A$^{\prime\prime}$ and B$^{\prime\prime}$,
respectively\footnote{However, since in Model B$^{\prime\prime}$,
$\tref= 0.01$ (\S \ref{sec:upar}), an amplitude of $A=0.03$ would
suffice in reproducing fluctuations at the level $\tasym \simeq 0.01 =
0.02 - \tref$ (see \S \ref{sec:steady}).}, which is unacceptably high
for O stars, as discussed in
\S~\ref{sec:intr}.

An interesting phenomenon occurs with variable ionizing sources is
that the mass of ionized gas shrinks as \tasym\ increases. For
instance, for $A=0.40$ in Model A$^{\prime\prime}$, the ionized mass
is reduced to 91\% of that of the initial steady-state equilibrium
model. The reason is simple if we recall that temperature fluctuations
(\pcite{peim67}) in the case of a process like recombination, which is
weighted towards lower temperatures ($\propto T^{-0.85}$), will lead
to a higher efficiency of the recombination rate, hence to a lower
mass of ionized gas in the nebula for a given mean ionizing flux. (The
time-averaged luminosity of all recombination lines, however, remains
the same.) Note finally the tendency of the mean temperature \tom\ to
increase, at large amplitudes, slightly above the equilibrium value,
as a result of the photoheating energy being absorbed more efficiently
during the IF phase.  The dotted line shows the approximate behavior
expected for the ionized mass using the function
$\left<M\right>/M_{eq} \simeq (1+ 0.5 \, \alpha
(\alpha-1)\tasym)^{1/\alpha} \tom/\teq$ with $\alpha = -0.85$ (adapted
from Equation 5 in \pcite{bin00}). This expression becomes invalid
above $\tasym > 0.2$.

\section{Discussion} \label{sec:disc}

\subsection{Intrinsic source variability} \label{sec:intr}

We have shown that variations of the ionizing luminosity can produce
appreciable temperature fluctuations across the nebula. However, a
careful analysis of our results leads us to believe that the proposed
mechanism for generating fluctuations is not viable. For instance, to
reproduce a $\tobs = 0.025$ as in the Orion nebula (\pcite{est98})
would require reproducing $\tasym = 0.015 = 0.025 - \tref$. This
occurs with Model B$^{\prime\prime}$ (more appropriate to Orion) when
$A=0.04$, that is, it requires a variability of $\pm 4$\% of the
ionizing luminosity of the exciting star \thetac\ (HD 37022), assuming
$\Pi \simeq 1$.  Such variability of $Q_H$ would imply variations of
order $\pm 250\,K$ in the stellar atmosphere (derived from \map\ using
$T_{\ast} = 39500$).  In the optical domain, the continuum variations
would be a lot less, about $\pm 1$\% in amplitude ($\Delta {\rm m}_{B}
= 0.02$ min-to-max). Furthermore, these estimates are based on a
variability timescale very close to the recombination timescale, which
is of order 20 years for Orion ($n_e \sim 5000$\,\cmc), otherwise
if it was shorter it would require variation amplitudes
a lot larger (c.f. Fig.~\ref{fig:perio}) in order to reproduce $\tobs=0.025$
(i.e. $\tasym = 0.015$).  Although \thetac\ has shown evidence of
variability in the stellar \heii\ absorption line (\pcite{con72};
\pcite{wal81}) and of a periodicity of 15 days in the \ha\ equivalent
width (\pcite{stahl}), no photometric variations of the continuum on
timescales from days to years have been reported (\pcite{van89}).

H\,{\sc ii} regions powered by Luminous Blue Variable (LBV) stars such
as the Carina nebula are potential candidates for finding a cause and
effect relation between \tsq\ and variability. One would expect \tsq\
to be significantly larger in nebulae excited by LBV
stars. Unfortunately, we have no information at hand about \tsq\ in
the Carina nebula for the purpose of comparison.
  
A general and compelling argument against source variability as the
main cause of the fluctuations is that for metal lines like \ciiiw,
\oiiiw, the bulk of their luminosity  occurs at a nebular depth such
that $y < 10^{-1.7}$, the fluctuations in temperature (due to
variability) are negligible in that inner zone of the nebula despite
substantial changes in $y$, as shown in Fig.~\ref{fig:struca}b. This
means that $\tasymo$ would be much smaller than $\tasymh$ calculated
with model B$^{\prime\prime}$ and $A=0.04$. (Even larger values of $A$
would lead to insignificant values of $\tasymo$.)  In this particular
model, it can be shown using \map\ that 90\% of \oiii\ flux is emitted
in the inner regions where no fluctuations occur due to
variability, as illustrated in Fig. \ref{fig:strucb}b.  Therefore our
models, based on the variability of $Q_H$, cannot realistically
reproduce the large observed \tobs\ deduced from intermediate
excitation emission lines.

\subsection{Variable shadows from moving opaque condensations} \label{sec:shad}

There are other mechanisms than variable stars that can produce a
transient variability in the propagating ionizing flux. For instance,
the displacement of dense neutral condensations in planetary nebulae or
\hii\ regions (proplyds) can introduce temporal variations in the
ionization structure. Within the shadow, behind these opaque
condensations, the gas is either neutral or partly ionized by the
diffused field from the surrounding nebula. If these condensations of
lateral size $D$ had a tangential velocity component $V$ (distinct
from the nebular gas), the tangential displacement of the shadow
inside the nebula would produce effects similar to that of an ionizing
source, which is switched off, then on again, generating in turns an RF
and IF, respectively.  The important timescales in this problem are
\trec\ and \tcros, the crossover timescale ($\tcros =D/V$) 
of the shadow over a distance given by the diameter of the opaque
eclipsing condensation.  We expect that the effect on the shadowed region would
be strongest when \tcros\ becomes comparable to \trec, since it would
first induce a strong temperature drop and substantial recombination
followed by the propagation of hot IF after the ``eclipse'' has
ended. On the global scale of the nebula, these shadowed regions would
cause temperature fluctuations similar to those calculated for a star
that is briefly turned off. Whether the temperature fluctuations
would be felt over the whole nebula would depend on how many such
shadowed regions fill the nebular volume.

\begin{table}  
   \setlength{\tabcolsep}{1em} 
 \setlength{\tabnotewidth}{0.8\columnwidth} 
  \tablecols{4} 
\begin{center}
\caption[]{Fluctuations from intermitently shadowed
ionized regions} 
\label{tab:shad}
\begin{tabular}{llll}             
            \noalign{\smallskip}\hline\hline\noalign{\smallskip}
 $\Pi$     & $\Delta$\tabnotemark{a}  & $\Pi \times \Delta$ & \tasym \\
            \noalign{\smallskip}
            \hline
            \noalign{\smallskip}
10 & 0.025 & 0.25 & 0.042 \\
5  & 0.05 & 0.25 & 0.081 \\
2.5  & 0.1 & 0.25 & 0.15 \\
1.25  & 0.2 & 0.25 & 0.23 \\
 & & & \\
10 & 0.1 & 1.0 & 0.063 \\
10 & 0.05 & 0.5 & 0.045 \\
10 & 0.025 & 0.25 & 0.042 \\
10 & 0.0125 & 0.125 & 0.021 \\
10 & 0.00625 & 0.0625 & 0.013 \\
\noalign{\smallskip}   
\hline\hline
\tabnotetext{a}{Fraction of cycle during which the ionizing
flux is turned off.}
\end{tabular}
\end{center}
   \end{table}

As an estimate of the \tsq\ expected in that case, we ran models
in which the source was turned off during a time $(\Pi \times \Delta)
\, \trec$ where $\Pi$ is the period of the off-on cycle in units of 
the recombination timescale and $\Delta$ is the fraction of the cycle
during which the source is off. The idea behind using periodicity to
approach this problem is that in this way we can crudely estimate the
nebular fluctuations due to the shadows by associating $\Delta$ to the
shadowed fraction, that is, to the effective covering factor of the
source due to all the neutral condensations present. To a first order,
this approach is validated by models for which we kept $\Pi \times
\Delta = 0.25$ constant, since we find that $\tasym \propto \Delta$,
at least when $\Delta $ is small (see first 4 models in
Table~\ref{tab:shad}).  A sequence of models in which we change $\Pi
\times \Delta$ is also given in Table~\ref{tab:shad}. Note that since
the amplitude of the variations are extreme in the case of an on-off
variability type, we expect that \tasymo\ would be substantial as
well\footnote{This is indirectly confirmed by the LINER model of
\scite{era95} who found a significantly higher \oiiitw/\oiiiw\
temperature sensitive ratio in their time-dependent model as a result
of the very large variability amplitudes of their source.}. For the
optimal regime where $\Pi \times \Delta \simeq 1$, a covering factor
of 4\% is required for the proplyds if we aim to reach $\tobs =
0.025$. Since these models were computed for a low value of \upar, we
can estimate that \tasym\ would be higher by factor about three, since
$\tasym \propto \sqrt{\upar}$ as discussed in \S \ref{sec:upar}. Hence
maybe a covering factor as low as 1\% would suffice.  However, the
fraction of the Orion nebula volume shadowed by proplyds is estimated
to be only 0.1\% (William Henney: private communication), which
therefore  rules out this mechanism.

\section{Conclusions}\label{sec:conc}

Our time-dependent models rule out the possibility that the ionizing
source variability be the cause of the variations because OB stars are
not known to vary at the required 15\% level (of $Q_H$). Furthermore,
even if they did, \tasymo\ would still be much smaller than our
computed \tasymh, making the observed values (c.f. \pcite{lur99};
\pcite{est98}) even more unattainable. As for the model of shadow crossings by
opaque condensations, it requires a space density of proplyds a factor
ten beyond that estimated by observations.

An interesting and general result about cyclicly variable ionizing
sources is that the nebula becomes ``on average'' less massive but
somewhat hotter. Despite the fluctuating size and temperature of the
nebula, the {\it time-averaged} luminosity of all recombination lines
remains the same as for the steady-state case. This is also the case
for the {\it time-averaged} energy radiated through all the forbidden
lines as compared to the steady-state case (individual forbidden line
ratios must in general come out different since the nebula is hotter).


\begin{acknowledgements}
The work of LB was supported by the CONACyT grant 32139-E.  PF
acknowledges support from the R\'egion Rh\^one-Alpes. The comments
received from William Henney about proplyds were particularly
useful. We thank Jane Arthur for her contribution during the workshop
on 3D-hydro in Mexico City (February 1999) during which the codes
\ygn\ and \ygz\ were written. We are also indebted to AR who
generously funded this event.
\end{acknowledgements}

\appendix{} \label{apx}

\section{\ygn\  --- The equations}
   \label{AppendixEquations} In this Appendix, we describe the set of
   equations used to follow R-type ionization fronts within a spherical
   nebula photoionized by a variable source. The new code, hereafter
   called \ygn, has been written in Fortran 77 and will be made
   accessible to researchers upon request.

We consider the problem of photoionization of a thick spherical gas shell by a central,
   time-varying source.  We make the following simplifying assumptions concerning the
   source and the gas distribution. 

\begin{itemize}
\item The ionizing radiation from the central source is taken to 
be monochromatic at a frequency \nubar\ $>$ \nuzero\ (see value in
Table~2) where $h$\nuzero $= 13.6$~eV is the
ionization potential of hydrogen.  The central source is considered
point-like (i.e. its characteristic size is much smaller than the
inner radius of the gas shell).
\item For the purpose of radiation transfer, the gas shell
consists only of hydrogen. The gas density is static in time and its
value is either a constant or a function of radius.
\item The atomic physics is simplified along the lines
developped in \ref{balance} and \ref{energy}. All essential physical
processes are considered (e.g. approximate cooling by metals), although
the estimation of their rates is limited to first order
approximations.
\end{itemize}
   
Given these assumptions, all variables of the problem depend only
on the radius from the central source (spherical symmetry) and on the
time.
   
   \subsection{Time-dependent transfer equation}
      The time-dependent equation of transfer for monochromatic ionizing
      radiation, and for spherically symmetric problem is (in spherical
      coordinates):
      \begin{equation}
	 \label{EquationTransfer}
	 \frac{1}{\mbox{\small c}} \frac{\partial {\cal F}}{\partial t}(r,t)
	 \;+\;
	 \frac{1}{r^2} \frac{\partial (r^2 {\cal F})}{\partial r}(r,t)
	 \;=\; - \kappa(r,t)
      \end{equation}
      where ${\cal F}$($r$,$t$) is the ionizing photon flux (in photon s$^{-1}$
      cm$^{-2}$) at a radius $r$ from the central source (in cm) and at the time $t$
      (in s), $\kappa(r,t)$ is the local opacity at \nubar\ (in photon s$^{-1}$
      cm$^{-3}$), and $c$ is the speed of light (in cm~s$^{-1}$). We
      have ignored the scattering of the ionizing radiation by dust and the
generation of the diffuse field,
      which would have introduced `local' source terms in equation
      (\ref{EquationTransfer}). 
      
      For a pure hydrogen medium and monochromatic radiation, the
      opacity yields the relation: \begin{equation}
      \label{EquationOpacity} \kappa(r,t) \;=\;
      \overline{a}(\overline{\nu}) \; \varepsilon \;
      \mbox{n}_{\mbox{\scriptsize H}}(r) \; \left[ 1 \,-\, f(r,t)
      \right] \; {\cal F}(r,t) \end{equation} where \abar(\nubar) is
      photoionization cross section of hydrogen at \nubar\ (in
      cm$^2$), $\varepsilon$ is the filling factor of the shell
      (assumed uniform and static), \nH($r$) is the hydrogen number
      density (in \cmc), and $f$($r$,$t$) is the ionization fraction
      of the hydrogen. The following relation has been used to compute
      \abar\ as a function of \nubar: 
\begin{equation}
      \label{EquationPhotoCrossSection}
      \overline{a}(\overline{\nu}) \; = \; 
       6.3 \times 10^{-18} 
      \; \left( \frac{\overline{\nu}}{\nu_0} \right)^{-3}
      \;\; \mbox{cm}^2
\end{equation} 
where the value of $\overline{\nu}$
      is chosen according to the type of ionizing source and is listed
      in Table~2. $\nu_0$ is the H ionizing
threshold. A monochromatic treatment
      implies that the progressive hardening of the ionizing radiation
      (and hence of the temperature) away from the central source
      cannot be taken into account. Furthermore, in the inner region
      of planetary nebulae and AGN, heating by photoionization of
      He$^+$ is substantial and lead to higher temperatures, this
      effect cannot 
      be taken into account in our simplified pure hydrogen nebula.

   \subsection{Time-dependent ionization balance of hydrogen} \label{balance}
      To solve the transfer equation (\ref{EquationTransfer}), we need to
      know the ionization fraction $f$($r$,$t$) of the hydrogen, and, therefore, to
      solve the ionization balance of hydrogen. The ionization fraction yields the
      following equation:
      \begin{equation}
	 \label{EquationIonizationBalance}
	 \begin{array}{ll}
	 \frac{\partial f}{\partial t}(r,t)
	 \;=\;
	 &
	 \underbrace{
	 \overline{a}(\overline{\nu}) \; 
	 \left[ 1 \,-\, f(r,t) \right]\;
	 {\cal F}(r,t)
	 }_{
	 \mbox{\scriptsize photoionization}
	 }
	 \\
	 &
	 +\;
	 \underbrace{
	 \gamma(\mbox{T}) \; 
	 \mbox{n}_{\mbox{\scriptsize H}}(r) \;
	 f(r,t) \, \left[ 1 \,-\, f(r,t) \right]
	 }_{
	 \mbox{\scriptsize collisional ionization}
	 }
	 \\
	 &
	 -\;
	 \underbrace{
	 \alpha_{\mbox{\scriptsize B}}(\mbox{T}) \;
	 \mbox{n}_{\mbox{\scriptsize H}}(r) \;
	 f^2(r,t) 
	 }_{
	 \mbox{\scriptsize recombination}
	 }
	 \end{array}
      \end{equation}
      where T($r$,$t$) is the temperature of the gas (in K), $\gamma$(T) is
      the collisional ionization coefficient (in s$^{-1}$~cm$^3$), and
      $\alpha_{\mbox{\scriptsize B}}$(T) is the  recombination rate to 
excited states of hydrogen (in
      s$^{-1}$~cm$^3$). We implicitly assume the on-the-spot
approximation in our calculations. The following analytical relations have been used to compute
      $\gamma$ and $\alpha_{\mbox{\scriptsize B}}$ as a function of T:
      \begin{equation}
	 \label{EquationColIonCooling}
	 \gamma(\mbox{T})
	 \;=\;
	 \exp\left[
	 \mbox{A} \,+\, \mbox{B} \times 
	 \left( \frac{\mbox{T}}{10^4 \, \mbox{K}} \right)^{-1}
	 \right]
	 \;\; \mbox{s}^{-1} \, \mbox{cm}^3      
      \end{equation}
      with A = $-$19 and B = $-$16, and
      \begin{equation}
	 \alpha_{\mbox{\scriptsize B}}(\mbox{T})
	 \;=\;
	 2.6 \times 10^{-13} \;
	 \left( \frac{\mbox{T}}{10^4 \, \mbox{K}} \right)^{-0.85}
	 \;\; \mbox{s}^{-1} \, \mbox{cm}^3
      \end{equation}
      
   \subsection{Time-dependent energy balance}  \label{energy}
      The third major equation describing our system is the energy balance equation
      for the gas:
      \begin{equation}
	 \label{EquationEnergyBalance}
	 \frac{\partial \mbox{P}}{\partial t}(r,t)
	 \;=\; -\frac{2}{3} \; \Lambda(r,t)
      \end{equation}
      where P is the gas pressure (in dyne cm$^{-2}$) and
      $\Lambda$ is the net cooling rate per unit of volume
      (in erg~s$^{-1}$~cm$^{-3}$). $\Lambda$ yields the
      relation:
      \begin{equation}
	 \label{EquationNetCooling}
	 \begin{array}{l}
	    \Lambda(r,t)
	    \;=\;
	    +\; 
	    \underbrace{
	    {\cal L}(\mbox{T}) \;
	    \mbox{n}_{\mbox{\scriptsize H}}^2(r) \; f^2(r,t)
	    }_{
	    \mbox{\scriptsize radiative cooling}
	    }
	    \\
	    \;\;\;+\;
	    \underbrace{
	    h\nu_0 \; \gamma(\mbox{T})\;
	    \mbox{n}_{\mbox{\scriptsize H}}^2(r) \; 
	    f(r,t) \, \left[ 1 \,-\, f(r,t) \right]
	    }_{
	    \mbox{\scriptsize cooling due to collisional ionization of H}
	    }
	    \\
	    \;\;\;+\;
	    \underbrace{
	    h\nu_0 \; q(\mbox{T})\;
	    \mbox{n}_{\mbox{\scriptsize H}}^2(r) \; 
	    f(r,t) \, \left[ 1 \,-\, f(r,t) \right]
	    }_{
	    \mbox{\scriptsize cooling due to collisional excitation of H}
	    }
	    \\
	    \;\;\;-\;
	    \underbrace{
	    \overline{a}(\overline{\nu}) \;
	    h(\overline{\nu} - \nu_0) \;
	    \mbox{n}_{\mbox{\scriptsize H}}(r) \;
	    \left[ 1 \,-\, f(r,t) \right] \;
	    {\cal F}(r,t)
	    }_{
	    \mbox{\scriptsize heating due to photoionization of H}
	    }
	 \end{array}
      \end{equation}
      where $h$ is the Planck constant (6.63 $\times$ \ten{-27} erg~s). The
      coefficients ${\cal L}$(T), $\gamma$(T), $q$(T) and \abar(\nubar) are given
      in Eq.~(\ref{EquationRadiativeCooling}), (\ref{EquationColIonCooling}),
      (\ref{EquationColExcCooling}) and (\ref{EquationPhotoCrossSection}),
      respectively.
      
      In order to have a realistic energy balance for the gas, the radiative cooling
      term includes the losses due to metals. ${\cal L}$(T)
      has been computed using the following relation, which is an analytical fit to
      the cooling function obtained for a photoionized gas using \map\  and assuming solar metallicities:
      \begin{equation}
	 {\cal L}(\mbox{T})
	 \;=\;
	 {\cal A} \;+\; {\cal B} \;
	 \left( \frac{\mbox{T}}{10^4 \, \mbox{K}} \right)^{2.5}
	 \;\; \mbox{erg s}^{-1}\mbox{ cm}^3
	 \label{EquationRadiativeCooling}
      \end{equation}
      where the values of ${\cal A}$, ${\cal B}$  depend on the type of
      environment. Three sets of values are listed in
      Table~2.
      
      \begin{table}\label{TableRadiativeCooling}
  \setlength{\tabcolsep}{0.2em} 
  \setlength{\tabnotewidth}{0.8\columnwidth} 
  \tablecols{3} 
\begin{center}
	 \caption{Parameters for the radiative cooling curve\tabnotemark{a} and the value of $\overline{\nu}$ }
	 \begin{tabular}{lccc}
	    \noalign{\smallskip}\hline\hline\noalign{\smallskip}
	    Environment & ${\cal A}$ & ${\cal B}$   &
$h{\overline{\nu}} $ (eV) \\
	    \noalign{\smallskip}\hline\noalign{\smallskip}
	    \hii\ region & 3.8 $\times$ \ten{-24} & 4.0 $\times$ \ten{-24} &
	     20\tabnotemark{b}\\
	    Planetary nebula & 4.0 $\times$ \ten{-24} & 3.5 $\times$ \ten{-24} &
	     40\tabnotemark{c}\\
	    Active nucleus & 4.2 $\times$ \ten{-24} & 3.0 $\times$ \ten{-24} &
	     43\tabnotemark{d}\\
	    \noalign{\smallskip}\hline\hline\noalign{\smallskip}
\tabnotetext{a} {${\cal A}$ and ${\cal B}$ are in
	 units of erg s$^{-1}$ cm$^3$.}
\tabnotetext{b}{Stellar atmosphere of temperature 40\,000\,K.}
\tabnotetext{c}{Black body of temperature 150\,000\,K.}
\tabnotetext{d}{Power law of index $-1.3$ truncated at 1\,keV.}
	 \end{tabular}
\end{center}
      \end{table}
      
      This radiative cooling term applies only to a fully ionized
      medium. We do not take into account how this term
      varies with the ionization parameter although we consider separately
      the cooling due to excitation or ionization of H$^0$, which can be
      significant across  ionization fronts.  To
      compute the {\it net} cooling rate we therefore added to
      Eq.~(\ref{EquationNetCooling}) the cooling terms due to both
      collisional ionization and collisional excitation of hydrogen as
      well as the heating term due to photoionization of hydrogen. The
      collisional excitation rate $q$(T) yields the following relation
      [similar to Eq.~(\ref{EquationColIonCooling})]:

      \begin{equation}
	 \label{EquationColExcCooling}
	 q(\mbox{T})
	 \;=\;
	 \exp\left[
	 \mbox{C} \,+\, \mbox{D} \times 
	 \left( \frac{\mbox{T}}{10^4 \, \mbox{K}} \right)^{-1}
	 \right]
	 \;\; \mbox{s}^{-1} \, \mbox{cm}^3 
      \end{equation}
      with C = $-$18 and D = $-$11. The determination of the constants
A, B, C and D are based on the coefficients found in \scite{ost89}.
      
   \subsection{Equation of state}
      In order to close our system of equations, we need to express the equation of
      state of the gas. We have used the perfect gas equation of state:
      \begin{equation}
	 \label{EquationPerfectGas}
	 \mbox{P}(r,t)
	 \;=\;
	 \left[ 1 + f(r,t) \right]\,
	 \mbox{n}_{\mbox{\scriptsize H}}(r)\,
	 k\,
	 \mbox{T}(r,t)
      \end{equation}
      where $k$ is the Boltzmann's constant (1.38 $\times$ \ten{-16} erg~K$^{-1}$).
      From Eq.~(\ref{EquationEnergyBalance}) and (\ref{EquationPerfectGas}), we can
      derive the equation for the temperature:
      \begin{equation}
	 \label{EquationTemperature}
	 \begin{array}{l}
	 \frac{\displaystyle \partial \mbox{T}}{\displaystyle \partial t}(r,t)
	 \;=\;
	 \frac{\displaystyle 1}{\displaystyle
	 k\,
	 \mbox{\footnotesize n}_{\mbox{\tiny H}}(r)\,
	 \left[ 1 + f(r,t) \right]
	 }\\
	 \\ 
	 \times\;
	 \left[
	 -\frac{\displaystyle 2}{\displaystyle 3} \; \Lambda_{\mbox{\scriptsize
	 net}}(r,t) \,-\,
	 \mbox{n}_{\mbox{\scriptsize H}}(r)\,
	 k\,
	 \mbox{T}(r,t)\,
	 \frac{\displaystyle \partial f}{\displaystyle \partial t}(r,t)
	 \right]
	 \end{array}
      \end{equation}
      
      Together, equations (\ref{EquationTransfer}), (\ref{EquationIonizationBalance})
      and (\ref{EquationTemperature}) form a closed system of equations.
      
\section{\ygn\ --- The algorithm} \label{sec:algo}
   \label{AppendixCode}
   In this Appendix, we describe the practical implementation of a code to solve the
   set of equations describe in \ref{AppendixEquations}
   
   \subsection{Boundary conditions}
      To solve the problem, we need to define a set of boundary conditions.
      First, we have to assume initial (i.e. at $t$ = 0~s) temperature and
      ionization structures for the whole shell (i.e. from its inner radius, \Rin, to
      its outer radius, \Rout), as well as the corresponding, initial distribution of
      ionizing radiation: 
      \begin{equation}
	 \label{EquationBoudary1}
	 \forall r,\;\; 
	 \mbox{R}_{\mbox{\scriptsize in}} \leq r \leq  \mbox{R}_{\mbox{\scriptsize
	 out}}\;\; 
	 \left\{
	 \begin{array}{lcl}
	    f(r,0) & = & f_0(r) \\
	    \mbox{T}(r,0) & = & \mbox{T}_0(r) \\
	    {\cal F}(r,0) & = & {\cal F}_0(r)
	 \end{array}
	 \right.
      \end{equation}
      
      Last, we need to know the ionizing photon flux of the source ${\cal
      F}_{\mbox{\scriptsize source}}$ at \Rin\
      as a function of time:
      \begin{equation}
	 \label{EquationBoundary2}
	 \forall t, \;\;
	 {\cal F}(\mbox{R}_{\mbox{\scriptsize in}},t)
	 \;=\;
	 {\cal F}_{\mbox{\scriptsize source}}(t)
      \end{equation}
      Assuming that the cavity inside the shell ($r$ $<$ \Rin) is empty, ${\cal
      F}_{\mbox{\scriptsize source}}$($t$) can be expressed as a
function  of the central source ionizing photons production rate, ${\cal
      Q}_{\mbox{\scriptsize source}}$ (in photon~s$^{-1}$):
      \begin{equation}
	 \label{EquationProductionRate}
	 {\cal F}_{\mbox{\scriptsize source}}(t)
	 \;=\;
	 \frac{
	 {\cal Q}_{\mbox{\scriptsize source}}(t)
	 }{
	 4\pi \, \mbox{\small R}_{\mbox{\tiny in}}^2
	 }
      \end{equation}

   \subsection{Algorithm} \label{sec:algot}
      
      In the following text, the indices $i$ and $j$ refer to the spatial
      and time axes, respectively.
      
      \subsubsection{Transfer equation} \label{sec:cour}
	 
\begin{figure}
	 \includegraphics[width=8cm, clip=true]{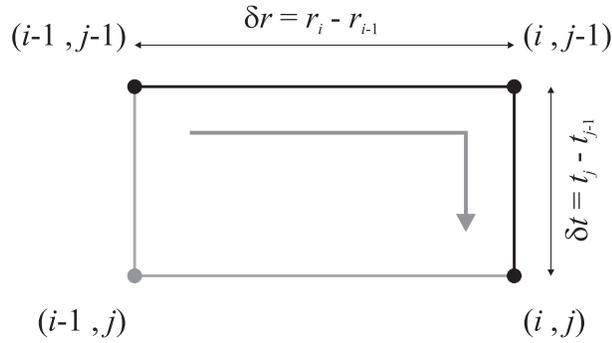}
	 \caption{Sketch of the progression of the computation along the
	 spatial ($i$ index) and temporal ($j$ index) grids.
	 \label{FigureAppendixStep}}
      \end{figure}

Once linearized, equation~(\ref{EquationTransfer}) allows us to
compute the ionizing photon flux ${\cal F}_{i,j}$ at a radius $r_i$
and time $t_j$, as a function of the ionization and temperature
structures of the shell and of the ionizing photon flux at radii
$r_{i-1}$ and $r_i$, and time $t_{j-1}$. A sketch of the progression
of the computation in the spatial and temporal grids is shown in
Fig.~\ref{FigureAppendixStep}. We have the following recursive
equation for ${\cal F}_{i,j}$:
\begin{equation}
  \begin{array}{ll}
  {\cal F}_{i,j}
  \;=\; &
  \underbrace{
  \begin{array}{ll}
  \frac{\mbox{c} \, \delta t}{\delta r} \, {\cal F}_{i-1,j-1} \,
  & \left(\frac{r_{i-1}}{r_i}\right)^2\\
  & \times \;
  \left(
  1 - \xi_{i-1,j-1} \, \delta r \, \frac{r_i^2}{r_{i-1}^2}
  \right)
  \end{array}
  }_{
\mbox{\scriptsize incoming photon flux (diluted and absorbed)}
}
	    \\
	    & \;+\;
	    \underbrace{
	    \left( 1 - \frac{\mbox{c} \delta t}{\delta r} \right ) \,
	    {\cal F}_{i,j-1}
	    }_{
	    \mbox{\scriptsize out-coming photon flux}
	    }
	    \\
	    \xi_{i-1,j-1} & \;=\;  
	    \overline{a}(\overline{\nu}) \, 
	    \varepsilon \,
	    \mbox{n}_{\mbox{\scriptsize H}}(r_{i-1}) \,
	    \left( 1 - f_{i-1,j-1} \right)
	    \end{array}
	 \end{equation}
where, $\delta r$ = $r_i - r_{i-1}$, $\delta t$ = $t_j - t_{j-1}$, and
c is the speed of light.  For this equation to be valid, we must
always have:

	 \begin{equation}
	    \label{EquationStepSize}
	    \begin{array}{l}
	       \mbox{c} \frac{\delta t}{\delta r} \;<\; 1\\
	       \delta r \;\ll\; \left( \frac{r_{i-1}}{r_i} \right)^2 \,
	       \xi_{i-1,j-1}^{-1}
	    \end{array}
	 \end{equation}
	 If $\delta t$ is larger than the typical atomic physic time
	 scales (e.g. the recombination time), the timestep is divided
	 in smaller time intervals. This prevents $\xi_{i-1,j-1}$
	 (i.e. the opacity) from changing significantly during the
	 timestep.
	 
      \subsubsection{Ionization balance} 
	 Once ${\cal F}_{i,j}$ is computed, the time-dependent ionization balance
	 [equation~(\ref{EquationIonizationBalance})] is then solved. The new
	 ionization fraction at $t_j$, $f_{i,j}$, is computed as a function of
	 $f_{i,j-1}$, T$_{i,j-1}$ and (${\cal F}_{i,j}$ + ${\cal F}_{i,j-1}$)/2. The
	 algorithm used to solve equation (\ref{EquationIonizationBalance}) is the
	 same as in \map\ (see Appendix in \pcite{bin85}).
	 
      \subsubsection{Temperature equation}
	 At last, the temperature T$_{i,j}$ is computed using the following recursive
	 equation [as derived from Eq.~(\ref{EquationTemperature})]:
	 \begin{equation}
	    \label{EquationTemperatureDiscrete}
	    \begin{array}{lll}
	       \mbox{T}_{i,j} & = & \mbox{T}_{i,j-1}
	       \;-\; \delta t \, \Theta_{i,j-1} 
	       \;-\; \Phi_{i,j-1}\\
	       \Theta_{i,j} & = & 
	       \frac{2}{3}
	       \frac{
	       \Lambda_{i,j}
	       }
	       {
	       k \,
	       \mbox{\small n}_{\mbox{\tiny H}}(r_i)\,
	       (1 + f_{i,j})
	       }\\
	       \Phi_{i,j} & = &
	       \frac{\mbox{\small T}_{i,j}}{1 + f_{i,j}} \,
	       (f_{i,j+1} - f_{i,j})
	    \end{array}
	 \end{equation}
	 $\Theta_{i,j}$ and $\Phi_{i,j}$ represent the change in temperature due to
	 the net cooling and the change in `molecular' weight, respectively. If
	 necessary, the ionization balance and the temperature equation were solved
	 iteratively.



\begin{thebibliography}{}

\bibitem[Baldwin et al.<1991>]{bf91} Baldwin, J.~A., 
Ferland, G.~J., Martin, P.~G., Corbin, M.~R., Cota, S.~A., Peterson, B.~M., 
\& Slettebak, A. 1991, \apj\ 374, 580 

\bibitem[Binette, Dopita \& Tuohy<1985>]{bin85} Binette, L., Dopita, M. A., \& Tuohy,
I. R. 1985,  \apj\ 297, 476

\bibitem[Binette \& Luridiana<2000>]{bin00} Binette, L. \& Luridiana, V. 2000, RevMexAA 36, 43


\bibitem[Binette, Luridiana \& Henney<2001>]{bin01} Binette, L., Luridiana, V. \& Henney, W. J. 2001,
RevMexAA Conf. Series 10, 19


\bibitem[Campbell<1988>]{camp88} Campbell, A.\ 1988, \apj\  335, 644 


\bibitem[Conti<1972>]{con72} Conti, P. 1972, \apjl\ 174, L79


\bibitem[Eracleous, Livio \& Binette<1995>]{era95} Eracleous, M., Livio, M. \& Binette, L. 1995,
\apjl\ 445, L1 

\bibitem[Esteban et al.<1998>]{est98} Esteban, C., Peimbert,   Torres-Peimbert,
S. \& Escalante, V. 1998, \mnras\ 295, 401


\bibitem[Ferruit et al.<1997>]{fer97} Ferruit, P., Binette, L., Sutherland, R. S., \&
P\'econtal, E. 1997,\aap\ 322, 73 

\bibitem[Kingdon \& Ferland<1995>]{king95} Kingdon, J. B. \& Ferland, G. J. 1995, \apj\ 450, 691

\bibitem[Kingdon \& Ferland<1998>]{king98} Kingdon, J. B. \& Ferland, G. J. 1998, \apj\ 506, 323

\bibitem[Liu et al.<2000>]{liu00} Liu, X.-W., Storey, P. J., Barlow, M. J.,
Danziger, I. J., Cohen, M., \& Bryce, M. 2000, MNRAS 312, 585

\bibitem[Luridiana et al.<1999>]{lur99} Luridiana, V., Peimbert, M., \& Leitherer,
C. 1999, \apj\ 527, 110

\bibitem[Luridiana, Cervi\~no \& Binette<2001>]{lur01} Luridiana, V., Cervi\~no, M., \& Binette,
L. 2001, \aap\ 379, 1017

\bibitem[Osterbrock<1989>]{ost89} Osterbrock, D.
1989, in Astrophysics of gaseous nebulae and active galactic nuclei,
University Science Books: Mill Valley

\bibitem[Peimbert<1967>]{peim67} Peimbert, M.  1967, \apj\ 150, 825 

\bibitem[Peimbert et al.<1995>]{peim95} Peimbert, M. Luridiana, V., \& Torres-Peimbert,
S. 1995,  RevMexAA 31, 131

\bibitem[P\'erez<1997>]{per97} P\'erez, E. 1997, MNRAS 290, 465




\bibitem[Stahl et al.<1993>]{stahl} Stahl, O., Wolf, B., G\"ang, T., Gummersbach,
C. A., Kaufer, A., Kovacs, J., Mandel, H., \& Szeifert, T. 1993, \aap\ 
274, L29

\bibitem[Stasi\'nska \& Szczerba<2001>]{sta01} Stasi\'nska, G. \& Szczerba, R. 2001,  
\aap\ 379, 1024 

\bibitem[Torres-Peimbert et al.<1990>]{tor90} Torres-Peimbert, S., Peimbert, M., \& Pe\~na,
M. 1990, \aap\ 233, 540

\bibitem[van Genderen et al.<1989>]{van89} van Genderen, A. M., et al. 1989, \aaps\ 79, 263

\bibitem[Walborn<1981>]{wal81} Walborn, N. R. 1981, \apjl\ 243, L37

\end{thebibliography}
\end{document}